# Semantic Oriented Intelligent Electronic Learning

Zeeshan Ahmed, Detlef Gerhard
*Mechanical Engineering Informatics and Virtual Product Development Division*
*Vienna University of Technology,*
*Phone: +43 1 58801 30726, Mobile: + 43 699 81302854,*
*Email: {zeeshan.ahmed, Detlef.gerhard}@tuwien.ac.at*
*Getreidemarkt 9/307 1060, Vienna, Austria*

## Abstract

*In this research paper we describe semantic oriented information engineering and knowledge management based solution towards E-Learning systems. We also try to justify the importance of proposed solution with respect to the E-Learning Approaches .i.e., Behavior, Objectivism, Cognitive and Construction. Moreover we briefly describe E-Learning, information engineering, knowledge management and some old and newly available technologies supporting development of E-Learning Systems in this research paper.*

## 1. Introduction

One of the major factor behind any student's progress and success in any subject is teaching effort, if the teacher is good and competent then the probability of getting much positive results from student can be very high. So we think it will be fine if we say,

"*Behind every successful student, there is a teacher*".

But the question is "What are/can be the major properties of a good and competent teacher?" In our opinion a good teacher should have four major properties .i.e.,
1. Good communication skills
2. Excellent knowledge and strong background of subject
3. Mature enough to understand student's problems
4. Skilled enough to answer the questions of students in best possible way.

Now a days the world has become globalized, technology especially computer science has progressed a lot, where computer science has contributed in almost every field of the world there it also has facilitated the field of education by providing the concept of Electronic Learning also renowned as E-Learning (EL) described in detail in Section 2. Where E-Learning is contributing in educating the people, there, it also facing some problems related to the major four properties of a good teacher .i.e., communication, analysis and search. To cope with these problems we have tried to support E-Learning with a new approach discussed in Section 5 based on an already proposed and underdevelopment approach called Semantic Oriented Agent Based Search (SOAS) [3, 4, 5], discussed in detail in Section 3 and justified its relativity to the field in Section 4.

## 2. Electronic Learning

Electronic Learning (E-Learning) is a computer based electronic educations system suitable for distance and flexible learning. E-Learning has four main approaches .i.e., Behavior, Objectivism, Congnitivism and Construction [6]. Behavior is based on stimulus and response mechanism, it provides communication facility between students and teacher: Using behavior student can ask questions from teacher, teacher can reply and teacher can also take some oral examination of students. But unfortunately there is no as such E-Learning system available which has full filled the requirements and successfully implemented Behavior. Objectivism based on the concept of transferring knowledge from E-Learning system to student e.g. power point slides, written texts, encoded movies etc. Congnitivism supports interaction with available information on E-Learning platform to interpret and build personal knowledge representation. Construction concentrates on the occurred changes in knowledge representation.

Depending upon above discussed four approaches of E-Learning, in our opinion currently E-Learning is trying to introduce and implement an intelligent E-Learning system which will be able to communicate with students, analyze their queries, search for the better and optimistic in time results, organize results in present and readable format and then respond back to the students as shown in Figure 1.

E-Learning is based on two main fields Information

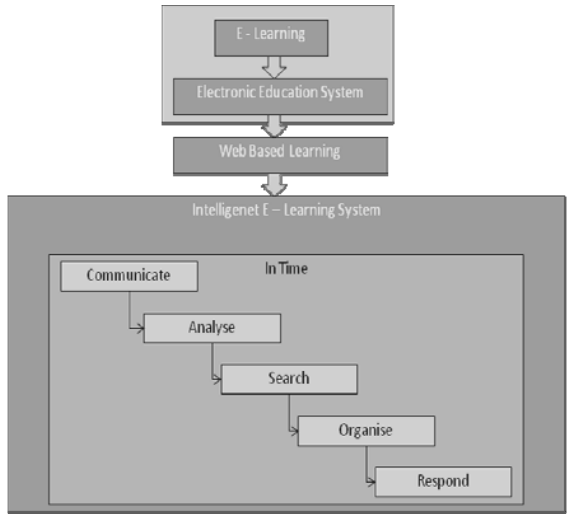

**Figure 1**.E-Learning [8]

Engineering (IE) and Knowledge Management (KM). IE provides a strategic approach for planning, design and development of Information Systems (IS) and KM provides the concepts of systematic and effective identification, utilization and exploitation of essential information. Moreover as shown in Figure.1 E-Learning depends on internet (World Wide Web) to become a successful and efficient distance based education system. Most of the E-Learning systems are developed and available on World Wide Web e.g. WebLessons, Learners Online, WebCurrents, Professional Services [7].

### 2.1. World Wide Web

World Wide Web is an automatic and cheapest global information sharing and communication system made up of three standards .i.e., Uniform Resource Identifier (URL), Hypertext Transfer Protocol (HTTP) and Hypertext Mark-up Language (HTML) to effectively store, communicate and share different forms of information. But still World Wide Web is struggling because the information is provided over the web in text, image, audio and video formats using HTML, considered unconventional in defining and formalizing the meaning of the context [1] which leads to the problem of communication, metadata based search, automatic excessive text preprocessing, cleaning raw data, automatically classifying documents into a taxonomy and browsing through large result lists with only relevant items.

### 2.2. Semantic Web and Web

To cope with the currently existing World Wide Web's problems and provide augmentation of meaningful contents in mark-up presentation over the web a semantic based solution "Semantic Web" was proposed [1]. The semantic web is an intelligent incarnation and advancement in World Wide Web to collect, manipulate and annotate information independently by providing effective access to the information. Semantic web provides categorization and uniform access to resources, promotes the transformation of World Wide Web in to semantically modeled knowledge representation systems and common framework which allows data to be shared and reused using Ontology [2].

### 2.3. Ontology and Semantic Web

Ontology is playing a vital role in solving the existing web problems by producing semantic aware solutions. Ontology makes machines capable of understanding the semantic languages that humans use and understand by producing the abstract modeled representation of already defined finite sets of terms and concepts involved in intelligent information integration and knowledge management [2]. Ontology can be developed using currently available and supported languages XML, RDF and OWL [2].

As the conclusive remarks based on already discussed Web, Semantic Web and Ontology in Section A, B and C, we can say that these technologies are not sufficient to support the development of intelligent E-Learning system. Currently available web based technologies can only facilitate by providing E-Learning based data on web for users to browse, access, interpret and download. We need a web based approach which can independently work to fulfill the major goals of intelligent E-Learning systems .i.e., independent communication, analysis, and search, organize and respond.

## 3. Semantic Web based Approach towards E-Learning

To contribute in the field of E-Learning by providing the concept of electronic efficient communications, user query analysis, intelligent search we have proposed an approach called Semantic Oriented Agent Based Search (SOAS) [3]. The proposed designed architecture of SOAS as shown in Figure 2. is consists of one Personal Agent (PA), Five dynamic processing units .i.e., Request Processing Unit (RPU), Agent Locator (AL), Agent Communicator (AC), List Builder (LB) and Result

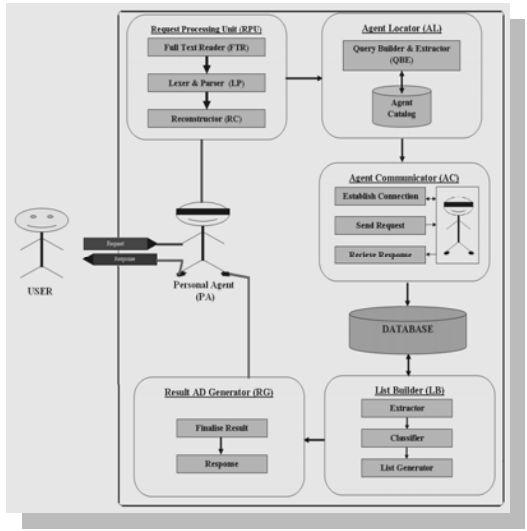

**Figure 2**. Semantic Oriented Agent based Search [3,4]

Generator (RG).

The architecture is designed in a way that the unstructured full text user request desiring some particular knowledge based information is given to PA. PA will simply transfer the received request to RPU, which will analyze and convert unstructured full text user request in to a structured semantic based data request. Constructed new semantic based search request will be passed and used by AL to find out the contact information of particular domain's available agents. Then using extracted contacts AC will contact, communicate, obtain information from available agents and stored in to the database. Stored information will be extracted and prioritized list of results will be generated by LB. Prioritized results will be forwarded to RG which will finalize the result by converting into the acceptable readable format and then will pass to PA. In the end PA will respond back to user by providing retrieved results. [3]

## 4. SOAS and E-Learning

Semantic Oriented Agent Based Search (SOAS) is capable of the supporting E-Learning system development which can be justified with respect to the approaches of E-Learning. Concluded from above discussion about SOAS in Section III that Personal Agent (PA) & Agent Communicator (AC) implements the concepts of Behavior and Objectivism, Agent Locator (AL) & List Builder (LB) are Cognitive and Result Generator (RG) is Constructive as described in Table 1.

| E-Learning Approach | Description | SOAS |
|---|---|---|
| Behavior | Stimulus– Response e.g. Question Answer | Personal Agent (PA) & Agent Communicator (AC) |
| Objectivism | Transfer of Knowledge e.g. Student to Teacher / Teacher to Student | Personal Agent (PA) |
| Congnitivism | Interact with information, Interpret, Build Personal Knowledge Representation | Agent Locator (AL) List Builder (LB) |
| Construction | Change knowledge representation | Result Generator (RG) |

**Table 1**. Justifications: SOAS and E-Learning

Moreover as we have already discussed that E-Learning based on two main fields Information Engineering and Knowledge Management, so, SOAS also contributes in information engineering by processing user's text based queries and locating desired information as well as organizing and managing information in to relation database management system.

## 5. Intelligent Electronic Teacher (IET)

Focusing on the targeted goals of E-Learning, faced problems in E-Learning system development, taking support of SOAS and based on available web technologies we propose a new approach called Intelligent Electronic Teacher (IET) [8] as shown in Figure 3.

We proposed IET to promote the idea of independency in E-Learning Systems. As we have already discussed above in Section II that there is a need of a web based approach for the intelligent E-Learning system development which can independently work and capable of doing the following jobs .i.e.,

- Independently communicate with students to listen and answer their questions and to take online oral examination.
- Analyze the natural language based student quires.
- Search for the best possible in time results.
- Classify and organize results.
- Respond back to the student in natural language.

As shown in Figure 3. Intelligent Electronic Teacher is promoting the development of Web Based E-Learning Systems for world globalised communication and taking the advantage of intelligently analyzing natural language based user quires by lexing, parsing and semantically modeling, intelligent search, classification and organization of data and reconstruction of the results in natural language.

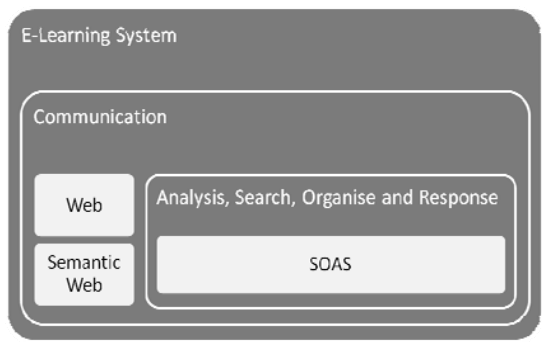

**Figure 3**. Intelligent Electronic Teacher (IET) [8]

## 6. Conclusion

In this research paper we have supported E-Learning with the already proposed and underdevelopment semantic oriented information engineering and knowledge management based approach. Moreover we have briefly described E-Learning, support of Web based technologies in E-Learning system development and contribution of Semantic Oriented Agent based Search towards E-Learning. In the end we have concluded with a new approach to contribute in web based E-Learning system development.